\documentclass[aps,preprintnumbers,showpacs,showkeys]{revtex4}
\usepackage{epsfig}
\usepackage{amssymb,amsmath,amsfonts,amsthm,graphicx}
\graphicspath{{./Figures/}}             

\newcommand{\noi}{\noindent}
\newcommand{\beq}{\begin{equation}}
\newcommand{\eeq}{\end{equation}}
\newcommand{\bea}{\begin{eqnarray}}
\newcommand{\eea}{\end{eqnarray}}

\newcommand{\ds}{\displaystyle}

\begin{document}

\title{Effects of dense quark matter on gluon propagators in lattice QC$_2$D}

\author{V.~G.~Bornyakov}
\affiliation{NRC ``Kurchatov Institute'' - IHEP,
142281 Protvino, Russia, \\
School of Biomedicine, Far East Federal University, 690950 Vladivostok, 
Russia}

\author{V.~V.~Braguta}
\affiliation{NRC ``Kurchatov Institute'' - ITEP, 117259 Moscow, Russia \\
School of Biomedicine, Far East Federal University, 690950 Vladivostok, Russia\\
Bogoliubov Laboratory of Theoretical Physics, Joint Institute for Nuclear Research, Dubna, 141980 Russia\\
Moscow Institute of Physics and Technology, Institutsky lane 9, Dolgoprudny, Moscow region, 141700 Russia }

\author{A.~A.~Nikolaev}
\affiliation{Department of Physics, College of Science, Swansea University, Swansea SA2 8PP, United Kingdom}

\author{R.~N.~Rogalyov}
\affiliation{NRC "Kurchatov Institute" - IHEP, 142281 Protvino, Russia}



\begin{abstract}
The transverse and longitudinal gluon propagators in the Landau gauge are studied in the two-color lattice QCD at nonzero quark chemical potential $\mu_q$.
Parameterization  of the momentum dependence of the propagators is provided for all values of chemical potential under study.
We find that the longitudinal propagator is infrared suppressed at nonzero $\mu_q$ with suppression increasing
with increasing $\mu_q$. The transverse propagator dependence on $\mu_q$ was found to be opposite: it is enhanced at large $\mu_q$. It is found, respectively, that the electric screening mass is increasing while the magnetic screening mass is decreasing with increasing $\mu_q$.
Nice agreement between the electric screening mass computed from the longitudinal propagator and the Debye mass computed earlier from the singlet static quark-antiquark potential was found.
We discuss how the dependence of the propagators on the chemical potential correlates with the respective dependence of the string tension.
Additionally, we consider the difference between two propagators as a function of the momentum and make interesting observations.
\end{abstract}

\maketitle


\section{Introduction}
\label{section0}

Understanding of the phase diagram of the strong interactions is of high importance for
experimental studies of hadronic matter created in relativistic heavy ion collisions.
The most difficult for theoretical investigation part of this phase diagram is at low temperature and high density.
Lattice QCD being the nonperturbative first principles approach is very successful at zero baryon density
but is inapplicable at high baryon density due to the so called sign problem \cite{Muroya:2003qs}.
This makes important to study the theories similar to QCD (QCD-like) but without 
sign problem. In particular, two popular QCD-like theories are 
QCD with $SU(2)$ gauge group \cite{Kogut:2000ek} (to be called below QC$_2$D) and QCD with 
nonzero isospin chemical potential \cite{Son:2000xc}. QCD
with the isospin chemical potential was intensively studied both within lattice and other approaches (see, for instance, \cite{Son:2000xc, Kogut:2002zg, Brandt:2017oyy, He:2005sp, Khunjua:2019lbv, Khunjua:2017mkc} ). In this paper we are going to focus on  QC$_2$D at nonzero quark chemical potential $\mu_q$.
Although two-color QCD  differs from three-color QCD, lattice study of QC$_2$D at nonzero quark chemical potential can provide us with important information about the properties of QCD with non-zero baryon density.

QC$_2$D  was studied using various approaches:
chiral perturbation theory \cite{Kogut:2000ek,Splittorff:2001fy,Kanazawa:2009ks},
Nambu-Jona-Lasinio model \cite{Brauner:2009gu,Sun:2007fc,He:2010nb}, quark-meson-diquark model
\cite{Strodthoff:2011tz,Strodthoff:2013cua}, random matrix theory
\cite{Vanderheyden:2001gx,Kanazawa:2011tt},  Dyson-Schwinger equations \cite{Contant:2019lwf}, massive perturbation
theory \cite{Kojo:2014vja,Suenaga:2019jjv}.
These studies suggested the following phase structure of low-temperature QC$_2$D. There is a hadronic phase
at $\mu_q < \mu_c = \mu_\pi/2$, Bose-Einstein condensation phase at $\mu_c < \mu_q < \mu_d$,
and the phase with diquark condensation due to the Bardeen-Cooper-Schrieffer mechanism
at $ \mu_q > \mu_d$.

It is worth to note that these approaches are also applicable to QCD at high baryon density. It is thus important to check them in the case of QC$_2$D confronting respective results with first principles lattice results.

Lattice studies of QC$_2$D were undertaken with both staggered fermions
\cite{Hands:1999md,Kogut:2001if,Kogut:2001na,Kogut:2002cm,Braguta:2016cpw,Bornyakov:2017txe,
Astrakhantsev:2018uzd,Wilhelm:2019fvp} for $N_f=4$ or, more recently, $N_f=2$  and Wilson fermions
\cite{Nakamura:1984uz,Hands:2006ve,Hands:2010gd,Hands:2011ye,Cotter:2012mb,Boz:2018crd,Iida:2019rah}
for $N_f=2$ mostly.  In general the lattice results supported the phase structure described above.

The question of the confinement-deconfinement transition in  QC$_2$D at low temperature is still under debate.
In our recent paper \cite{Bornyakov:2017txe} we studied $N_f=2$ lattice QC$_2$D with staggered fermionic action at
high quark density and $T=0$ and demonstrated that the string tension $\sigma$ decreases
with increasing $\mu_q$ and becomes compatible with zero for $\mu_q$ above 850~MeV.
The simulations were carried out at small lattice spacing $a = 0.044$ fm which was
few times smaller than in all other lattice studies. 
This allowed to reach the domain of large quark chemical potentials avoiding strong lattice artifacts.
In a more recent paper \cite{Boz:2019enj}, where $N_f=2$ lattice QC$_2$D with Wilson fermionic action
was studied, the authors did not find the confinement-deconfinement transition at low temperature. It is worth to note
that in  \cite{Boz:2019enj}  rather coarse lattices were used with lattice spacings three times or more larger than
in our study \cite{Bornyakov:2017txe}. Thus the range of large $\mu_q$ where we found the transition to deconfinement was
reached in \cite{Boz:2019enj} at parameter $a \mu_q > 0.5$ implying possibility of strong lattice artifacts.

In this paper we concentrate on the study of the  Landau gauge gluon propagators in $N_f=2$ lattice QC$_2$D at zero temperature
and varying quark chemical potential. We use the same lattice action as in  \cite{Bornyakov:2017txe, Astrakhantsev:2018uzd}
and in fact the same set of lattice configurations. Our goal is to study how the gluon propagators change when
QC$_2$D goes through its transitions mentioned above: from hadron phase to superfluid phase, confinement-deconfinement transition, disappearance of the spatial string tension.
Some results of our study of the gluon propagators were presented  in \cite{Bornyakov:2019jfz}.
Here we extend the range of $\mu_q$ values, make more detailed comparison of two definitions of the screening masses
and consider in more detail the momentum dependence of the
gluon propagators. We also study a new observable, the difference between the (color-)electric and magnetic
propagators and study its dependence on the momentum and quark chemical potential.

The gluon propagators are among important quantities to study, e.g. they play crucial role in the Dyson-Schwinger equations approach.
Landau gauge gluon propagators in non-Abelian gauge theories at zero and nonzero temperature were
extensively studied in the infrared range of momenta by various methods. We shall note lattice gauge theory,
Dyson-Schwinger equations, Gribov-Zwanziger approach. At the same time the studies in the particular case of
nonzero quark chemical potential are restricted to a few papers only. For the lattice QCD this is explained by the sign problem mentioned above.

The gluon propagators in lattice QC$_2$D at zero and nonzero $\mu_q$ were studied for the first time in
\cite{Hands:2006ve}. This study was continued in  \cite{Boz:2013rca,Hajizadeh:2017ewa,Boz:2018crd}.
The main conclusion of Ref.~\cite{Boz:2018crd} was that the gluon propagators practically do not
change for the range $\mu_q < 1.1$~GeV.  Our main conclusion is opposite.
We found substantial influence of the quark chemical potential on the gluon propagators starting from rather
low values ($\mu_q \sim 300$ MeV) and increasing with increasing $\mu_q$. Part of our results were presented in
\cite{Bornyakov:2019jfz}.
The gluon propagators in QC$_2$D at nonzero $\mu_q$ were also studied in Ref.~\cite{Contant:2019lwf} with  help
of the Dyson-Schwinger equations approach and in  Ref.~\cite{Suenaga:2019jjv} using the
massive Yang-Mills theory approach at one-loop. The authors emphasize that after the agreement with the lattice results
for the gluon propagators will be reached their methods could be applied to real QCD at nonzero baryon density. Thus to provide unbiased lattice results is very important.

The paper is organized as follows. In Section~\ref{section1} we specify details of the lattice setup to be used: lattice action, definition of the propagators and details of the simulation.
In the next  Section we  present 
the numerical results for the momentum dependence of the propagators and our fits to the data.
Section~\ref{section3} is devoted to the screening masses computation and study of their dependence on the chemical potential. In Section~\ref{section4} results for the difference between the longitudinal and the transverse propagators are presented.
The last section is devoted to the discussion of the results and to conclusions to be drawn.

\section{Simulation details}
\label{section1}

We carry out our study using $32^4$ lattices for a set of the chemical potentials in the range $a\mu_q \in (0, 0.5)$.
The tree level improved Symanzik gauge action~\cite{Weisz:1982zw}
and the staggered fermion action with a diquark source term~\cite{Hands:1999md} were used.
The  lattice configurations were generated at a small  value of the diquark source term coupling
$\lambda=0.00075$ which was much smaller than the quark mass in lattice units $am_q=0.0075$.
More details on the generation of these lattice configurations can be found in Ref.~\cite{Bornyakov:2017txe}.
The pion mass for this ensemble is rather large,  $m_{\pi}/\sqrt{\sigma}=1.56(8)$.
In this paper we prefer to use the dimensionless quantities of the type $m^2/\sigma$ using the
value $\sqrt{\sigma} a = 0.106(1)$ \cite{Bornyakov:2017txe}
 for this purpose. In case we use the physical units the value for the
 Sommer scale $r_0=0.468(4)$~fm ~\cite{Bazavov:2011nk} and relation $r_0/a=10.6(2)$
 \cite{Bornyakov:2017txe}
 are used to convert the lattice spacing $a$ into physical units.

  To reach high quark densities without lattice artifacts one needs sufficiently small lattice spacing 
to satisfy condition $a\mu_q \ll 1$. At the same time, to study the gluon propagators in the infrared region it is 
necessary to employ large physical volume. 
As a result of a compromise between these two requirements our lattice size is rather moderate: $L = 3.4/\sqrt{\sigma} = 1.4$~fm.   
This implies a potential problem of large finite volume effects at small momenta. 
We come to this problem again at the end of this section.

In the Introduction we briefly described the phase diagram of dense QC$_2$D at zero temperature. 
Here we want to transcribe the boundaries of  this phase diagram in units of $\sqrt{\sigma}$ using results
obtained in our previous papers \cite{Braguta:2016cpw, Bornyakov:2017txe, Astrakhantsev:2018uzd}. 
For small values of the chemical potential $\mu_q  <
\mu_c $, where $\mu_c=m_\pi/2 \approx 0.78 \cdot \sqrt{\sigma}$,  the system is in the hadronic phase. 
In this phase the system exhibits confinement and chiral symmetry is broken. 
At $\mu_q  = \mu_c $  there is a second order
phase transition to a phase where scalar diquarks form a Bose-Einstein condensate (BEC phase). Enhancing the baryon density further, we proceed to dense matter. At sufficiently high baryon density 
some observables of the system under study can be determined   using Bardeen-Cooper-Schrieffer theory (BCS phase).
In particular, the baryon density is well described by the density of noninteracting fermions which occupy a Fermi sphere of radius $r_F = \mu_q$. The diquark condensate, which plays the role of a condensate of Cooper pairs, is proportional to the Fermi surface.

In addition to the transition to the BCS phase  we found \cite{Bornyakov:2017txe} the confinement-deconfinement transition  at $\mu_q/\sqrt{\sigma}  \sim 2.1$. This transition
manifests itself in a rise of the Polyakov loop and vanishing of the string tension. It is interesting that  the transition
to the BEC phase and the confinement-deconfinement transition are located close to each other as show our preliminary results.
It was also observed in \cite{Bornyakov:2017txe} that above the deconfinement transition the spatial string tension $\sigma_s$  monotonously decreases and vanishes  at  $\mu_q/\sqrt{\sigma}  \sim 4.2$.

In our study of the gluon propagators we employ the standard definition of the lattice gauge vector
potential $A_{x,\mu}$ \cite{Mandula:1987rh}:
\beq
A_{x,\mu} = \frac{1}{2iag}~\Bigl( U_{x\mu}-U_{x\mu}^{\dagger}\Bigr)
\equiv A_{x,\mu}^a \frac{\sigma_a}{2} \,.
\label{eq:a_field}
\eeq
The lattice Landau gauge fixing condition is
\beq
(\nabla^B A)_{x} \equiv {1\over a} \sum_{\mu=1}^4 \left( A_{x,\mu}
- A_{x-a\hat{\mu},\mu} \right)  = 0 \; ,
\label{eq:diff_gaugecondition}
\eeq

\noi which is equivalent to finding an extremum of the gauge-fixing functional

\beq
F_U(\omega)\; =\; \frac{1}{4V} \sum_{x\mu}\ \frac{1}{2} \ T\!r\; U^{\omega}_{x\mu} \;,
\label{eq:gaugefunctional}
\eeq

\noi with respect to gauge transformations $\omega_x~$.
To fix the Landau gauge we use the simulated annealing (SA) algorithm
with finalizing overrelaxation \cite{Bornyakov:2009ug}.
To estimate the Gribov copy effect,
we employ five gauge copies of each configuration; however, the difference between the "best-copy" and "worst-copy" values of each quantity under consideration lies within statistical errors.

The gluon propagator $D_{\mu\nu}^{ab}(p)$ is defined
as follows:
\beq
D_{\mu\nu}^{ab}(p) = \frac{1}{Va^4}
    \langle \widetilde{A}_{\mu}^a(q) \widetilde{A}_{\nu}^b(-q) \rangle\;,
\qquad
\eeq
where
\beq
\widetilde A_\mu^b(q) = a^4 \sum_{x} A_{x,\mu}^b
\exp\Big(\ iq(x+{\hat \mu a\over 2}) \Big),
\label{eq:gluonpropagator}
\eeq
$q_i \in (-N_s/2,N_s/2]$, $ q_4 \in (-N_t/2,N_t/2]$  and
the physical momenta $p_\mu$ are defined by the relations $ap_{i}=2 \sin{(\pi q_i/N_s)}$,
$ap_{4}=2\sin{(\pi q_4/N_t)}$.

At nonzero $\mu_q$ the $O(4)$ symmetry is broken and
there are two tensor structures for the gluon propagator \cite{Kapusta:2006pm}~:
\beq
D_{\mu\nu}^{ab}(p)=\delta_{ab} \left( P^T_{\mu\nu}(p)D_{T}(p) +
P^L_{\mu\nu}(p)D_{L}(p)\right)\,.
\eeq
We consider the soft modes $p_4=0$ and use the notation $D_{L,T}(p)=D_{L,T}(0,|\vec{p}|)$.

Next we come back to discussion of the finite volume effects. At sufficiently high density the chromoelectric screening length determined as the inverse of the chromoelectric mass is 
estimated in perturbation theory as follows: 
$$ l_E = {1 \over  m_E} \sim {1 \over  g(\mu_q) \mu_q} $$
Our results are in agreement with this prediction as will be demonstrated in
Section~\ref{section3}. Thus we expect that for sufficiently large $\mu_q$ there should be no large finite 
volume effects for the longitudinal propagator $D_L(p)$. 

The screening length associated with the transverse propagator $D_T(p)$ is defined as the inverse of the chromomagnetic screening mass $m_M$. Perturbation theory predicts zero value of the magnetic screening mass at large chemical potentials \cite{Son:1998uk};
for this reason, the nonperturbative estimates of $m_M$ are
of particular interest.

Perturbation theory gives some evidence that, at sufficiently large $\mu_q$, the chromomagnetic screening mass goes down, the respective screening length becomes large, and to study the infrared behavior of $D_T(p)$ large lattices are needed. 
It should be noticed that these arguments apply 
to QCD at high baryon density as well.   

\section{Momentum dependence}
\label{section2}
In this section we consider the momentum dependence of 
the gluon propagators for various values of $\mu_q$.
The propagators are renormalized according to the MOM 
scheme to satisfy the condition

\beq
D_{L,T}(p=\kappa) = 1/\kappa^2
\eeq
at $\kappa= 12.6 \sqrt{\sigma}$.

\begin{figure}[tbh]
\hspace*{-10mm}
\includegraphics[width=10cm,height=8cm,angle=0]{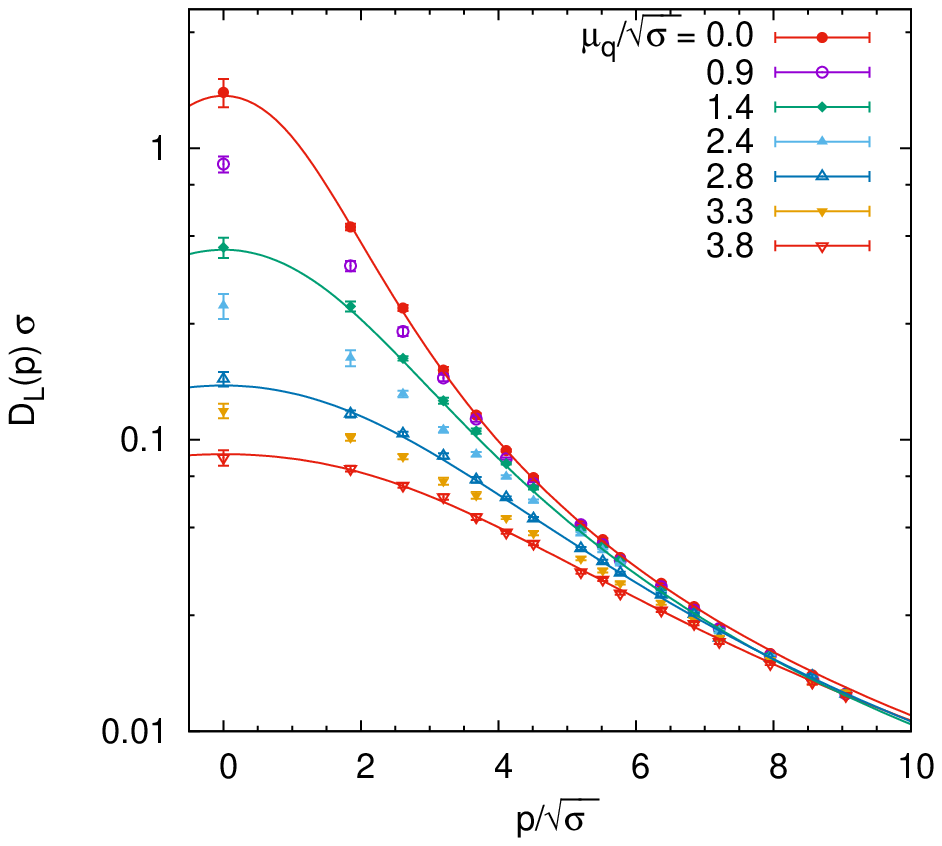}
\hspace*{-25mm}
\includegraphics[width=10cm,height=8cm,angle=0]{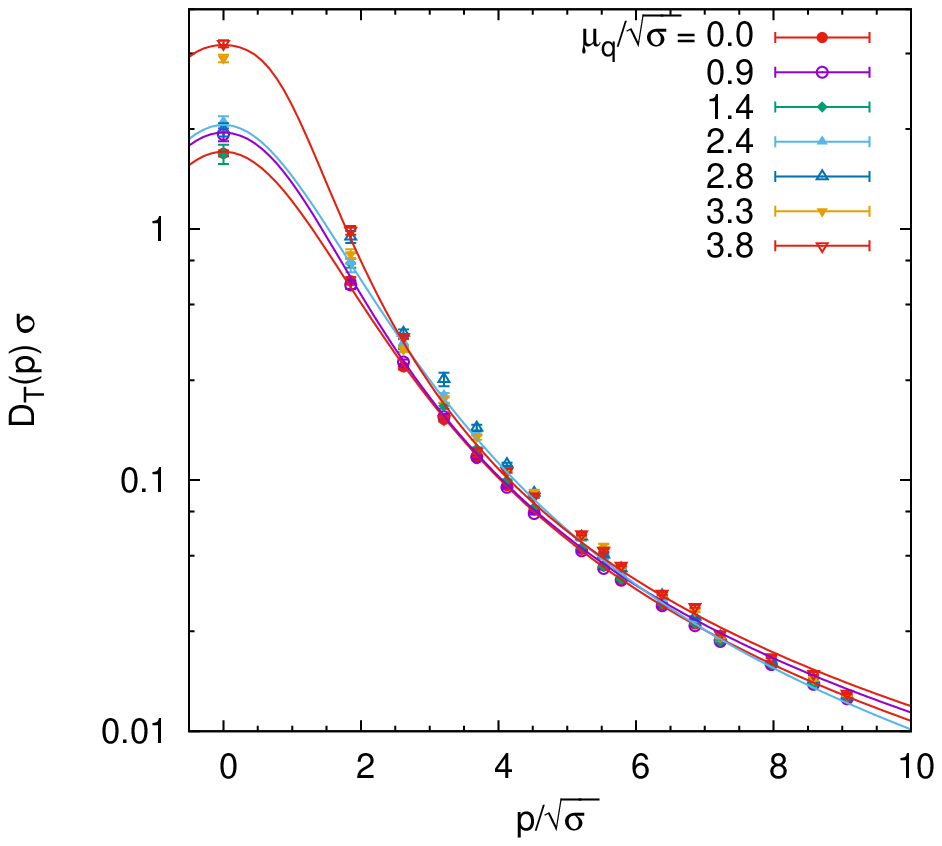}
\caption{The propagators $D_L$ (left) and $D_T$ (right) as functions of $p$
at different values of $\mu_q$. The curves show results of the fit to
eq.~(\ref{eq:GS_fit_widely_used})}
\label{fig:DLT_vs_p}
\end{figure}

In Fig.\ref{fig:DLT_vs_p}(left) we present the momentum dependence for the longitudinal propagator
$D_L(p)$ for seven selected values of $\mu_q$.
One can see that the infrared suppression of the propagator is clearly increasing with increasing $\mu_q$.
This infrared suppression hints on
the increasing of the electric screening mass. We will study the screening mass in the next section.
The increasing of the infrared suppression of $D_L(p)$ with increasing $\mu_q$ is analogous
to the well established behavior of $D_L(p)$ with increasing temperature in the deconfinement phase of both gluodynamics and QCD.

In Fig.\ref{fig:DLT_vs_p}(right) the momentum dependence for the transverse propagator $D_T(p)$ for the
same values of $\mu_q$ is shown. It is clear that $D_T(p)$ is much less sensitive to changes of $\mu_q$.
We found decreasing of the respective screening mass at large $\mu_q$ as will be discussed in the next section.
It is known that at a finite temperature the propagator  $D_T(p)$ has a clear maximum at the value of momentum increasing with temperature. Our data 
give no evidence for such maximum at a small momentum, however, we cannot exclude its existence.

We would like to provide an interpolation function for our data.
It was demonstrated  many times
\cite{Dudal:2010tf,Cucchieri:2011ig,Oliveira:2012eh,Dudal:2018cli,Aouane:2011fv}
that the infrared behavior of the gluon propagators
at zero and finite temperature can be well described by the fit function  which is
the tree level prediction of the Refined Gribov-Zwanziger approach, \cite{Dudal:2008sp}
\beq\label{eq:GS_fit_widely_used}
D_{L,T}(p)= Z_{L,T}\;{1 + \delta_{L,T} \, p^2 \over p^4+2 R_{L,T} \, p^2 + M_{L,T}^2 }\; .
\eeq
Our data for nonzero momentum start at rather large value
$p_{min}/\sqrt{\sigma} = 1.85(2)$. For this reason, the results of our fit over the infrared region may suffer from finite volume effects. Still, we believe that our results provide qualitatively correct dependence on $\mu_q$, in particular for $D_L$ at large $\mu_q$, see relevant discussion in  Section~\ref{section2}.

We found \cite{Bornyakov:2019jfz} that the fit of the data based on a 
one-loop perturbative expression
works well for $p>p_{cut}$, where 
\beq\label{eq:fit_domain_bound}
p_{cut} = 3.8 \sqrt{\sigma} + \mu_q\ .
\eeq
for $D_L$ and $p_{cut} = 6.0~\sqrt{\sigma}$ for $D_T$.
We perform the fit  (\ref{eq:GS_fit_widely_used}) over the domain $p < p_{cut}$; extending the fitting range above $p_{cut}$ results in a 
substantial decrease of the fit quality in most cases.

The results for the fit parameters for $D_L(p)$ are presented in Appendix A, Table~\ref{tab:LOW_MOM_fit_fun_L}.
The fits for large $\mu_q$ were not successful.
Using the Table the practitioners of other approaches to 
QC$_2$D can compare their results with ours\footnote{We
do not present correlations between the fitting parameters
$M, R,$ and $\delta$.  
For this reason the error in $D_{L,T}(p)$ evaluated from the data in Tables~1 and~2 assuming zero correlations between parameters provides an overestimated error. In so doing, 
the error in $Z$ can be safely neglected and we do not show it.}.

In practice we fitted the ratio $D_{L,T}(p)/D_{L,T}(p_0)$ with
$p_0/\sqrt{\sigma}=6.3$. This allowed us to decrease uncertainties in the fit parameters
$M_{L,T}^2, R_{L,T}, \delta_{L,T}$. Respectively, the parameters $Z_{L,T}$  were not determined from the fitting procedure
but recomputed (for renormalized propagator) via the relation
\beq
Z_{L,T} = D_{L,T}(p_0) \,{ p_0^4+2 R_{ L,T} \, p_0^2 + M_{L,T}^2   \over 1 + \delta_{L,T} \, p_0^2 }\; .
\eeq
Results of the fits for $D_L(p)$ are also shown in  Fig.\ref{fig:DLT_vs_p}(left) together with the lattice data.
In the hadron phase, the propagators change
insignificantly with increasing $\mu_q$.  For this
reason, absence of a systematic dependence of the parameters on $\mu_q$ at small $\mu_q$ is
not a surprise. Beyond the hadron phase, the parameters $M_L^2$, $R_L$, and $1/\delta_L$ show a similar behavior: they increase  with $\mu_q$.

In the case of the transverse propagator the fits were successful for $\mu_q / \sqrt{\sigma} < 3.0$, see
Table~\ref{tab:LOW_MOM_fit_fun_T}. The fit parameters
$M_T^2$, $R_T$  and $1/\delta_T$ again show qualitatively
similar dependence on $\mu_q$. Their values are lower at
the intermediate values $ 1.0 <  \mu_q/\sqrt{\sigma} < 1.8$ than in the hadron phase
and then increase again at $\mu_q/\sqrt{\sigma} \gtrsim 1.8$ to roughly the same values ($M_T^2$ and $R_T$) or to higher values  ($1/\delta_T$) than in the hadron phase.

\begin{figure}[tbh]
\hspace*{-5mm}
\includegraphics[width=9cm]{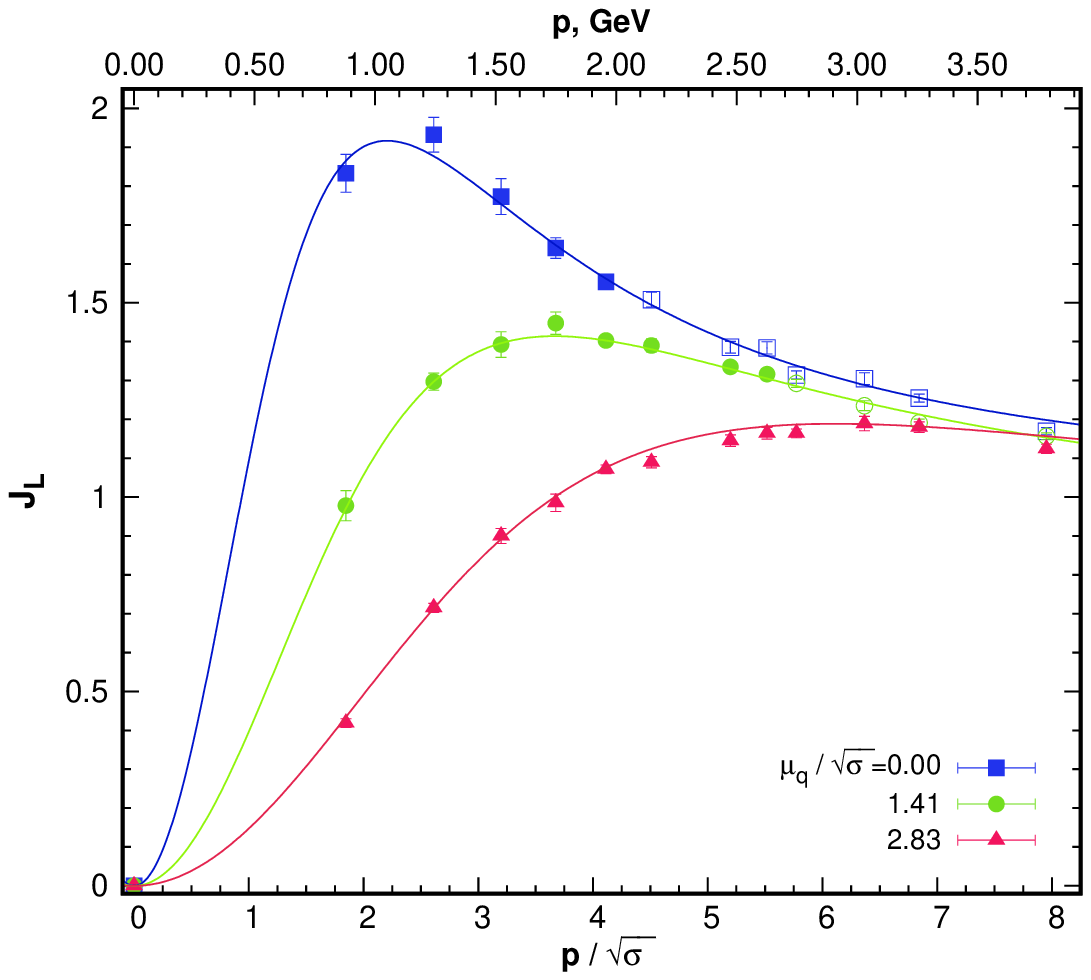}
\hspace*{-15mm}
\includegraphics[width=9cm]{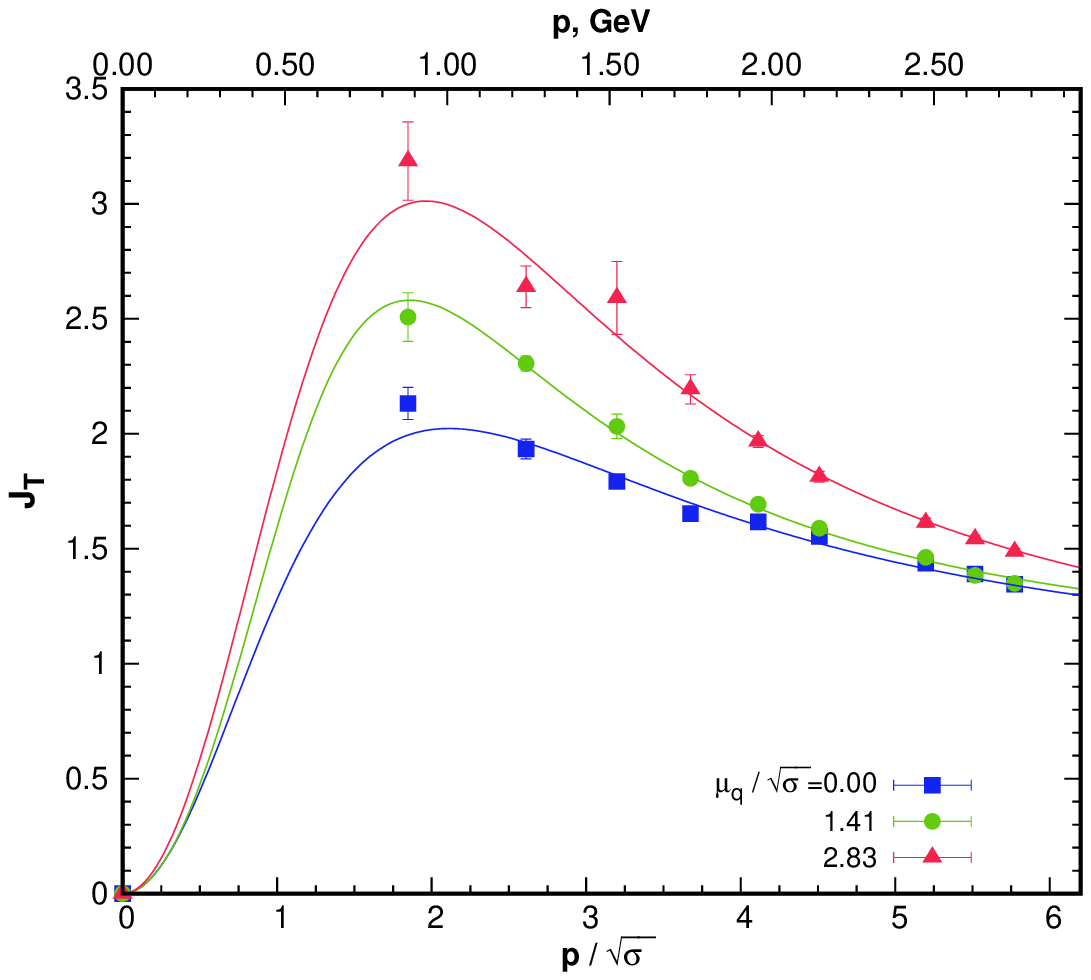}
\vspace*{0mm}
\caption{Dressing functions $J_L$ and $J_T$ as functions of $p$ at different values of $\mu_q$.
Empty symbols in the left panel are those beyond our fitting range (\ref{eq:fit_domain_bound}).
}
\label{fig:D0_vs_mu}
\end{figure}

It is instructive to look also at the respective dressing functions $J_{L,T}(p)$ defined  as
\beq
J_{L,T}(p) = p^2 \, D_{L,T}(p)
\eeq
It is seen in Fig.\ref{fig:D0_vs_mu} (left) that with increasing  $\mu_q$
the maximum of the longitudinal dressing function goes down and shifts to the right, thus approaching dressing function of a massive scalar particle. We note once more that this dependence on
$\mu_q$ is very similar to dependence on the temperature, see e.g. Ref.~\cite{Fischer:2010fx}.

 As can be seen in Fig.~\ref{fig:D0_vs_mu}(right) the transverse dressing function shows instead infrared enhancement with increasing $\mu_q$.
This is in agreement with the disappearance of 
the magnetic field screening at extremely large 
quark chemical potential predicted in \cite{Son:1998uk}.


\section{Screening masses}
\label{section3}

The widely used definition of the screening mass, see the review \cite{Maas:2011se} and references
therein, is through the inverse of the propagator at zero momentum
\beq
\label{maas}
m_E^2={1\over D_L(0)}\;, \qquad  \; m_M^2={1\over D_T(0)}.
\eeq
It is clear, that the screening mass defined by eq.~(\ref{maas}) depends on renormalization. Moreover, it is rather sensitive to the finite volume effects.
Loosely speaking, eq.~(\ref{maas})  characterizes ``the total amount'' of the interaction since
\beq
{1\over m_{E,M}^2} = \int dx_4 d\vec x D_{L,T}(x_4, \vec x),
\eeq
where $D_{L,T}(x_4,\vec x)$ are the propagators in the coordinate representation.

We also consider another definition of the screening mass using fitting of $D_{L,T}^{-1}(p)$ at low momenta by
Taylor expansion in $p^2$:
\beq\label{eq:LOW_MOM_fit_fun}
D_{L,T}^{-1}(p) = Z ^{-1}\large( \tilde{m}_{E,M}^2 + p^2 + c_4 \cdot (p^2)^2 +... \large)\; .
\eeq
This method was used in \cite{Bornyakov:2011jm} in the  studies of lattice QCD at finite temperatures 
and we applied it to QC$_2$D in \cite{Bornyakov:2019jfz}. In fact it would be more consistent to use the Yukawa  type fitting  function
\beq\label{scrmass_mom1}
D_{L,T}^{-1}(p) = Z ^{-1}(\tilde{m}_{E,M}^2 + p^2)
\eeq
as was done in \cite{Bornyakov:2010nc,Oliveira:2010xc,Silva:2013maa} in the studies of lattice gluodynamics at zero and finite temperatures.
It was shown in \cite{Silva:2013maa} that the  Yukawa type function (\ref{scrmass_mom1}) provides a 
constant value for $\tilde{m}_{E}^2$ over rather wide range of momenta in the infrared.
The reason we are using function  (\ref{eq:LOW_MOM_fit_fun}) rather than function (\ref{scrmass_mom1}) is that
we have no enough data points in the infrared region where the propagator can be described by the function
(\ref{scrmass_mom1}). Thus, to obtain a reasonable fit results we have to use terms up to $(p^2)^2$ for $D_L(p)$ and
terms up to $(p^2)^3$ for $D_T(p)$. Still, we hope that 
making use of the fit function (\ref{eq:LOW_MOM_fit_fun})
provides reasonably good estimates of the parameters in eq.~(\ref{scrmass_mom1}).

Let us note that the  definition of $\tilde{m}_{E,M}^2$ can be related to the definition of the
correlation length:
\beq
\label{mass_def1}
\tilde{m}_{E,M}^2=\xi_{E,M}^{-2},
\eeq
where the  correlation length $\xi_{E,M}$ is conventionally defined   in terms of the correlation function (propagator in our case)  by the expression \cite{ShangKengMa}
\beq
\label{eq:correlation}
 \xi^2 = \frac{1}{2} {\int_V dx_4 d\vec x  \tilde D(x_4, \vec x) |\vec x|^2
\over  \int_V dx_4 d\vec x   D(x_4, \vec x)} =
- {1\over 2 D(0, \vec 0)}\; \sum_{i=1}^3
\left({d\over dp_i}\right)^2\Big|_{\vec p=0} D(0, \vec p)\; .
\eeq

Even after the propagators are renormalized
the definitions of the screening mass (\ref{mass_def1}-\ref{eq:correlation}) and (\ref{maas}) differ in general by a factor
which may depend on the chemical potential or temperature. Its temperature dependence was found in
$SU(3)$ gluodynamics  \cite{Silva:2013maa}.

\begin{figure}[tbh]
\hspace*{-5mm}
\includegraphics[width=9cm]{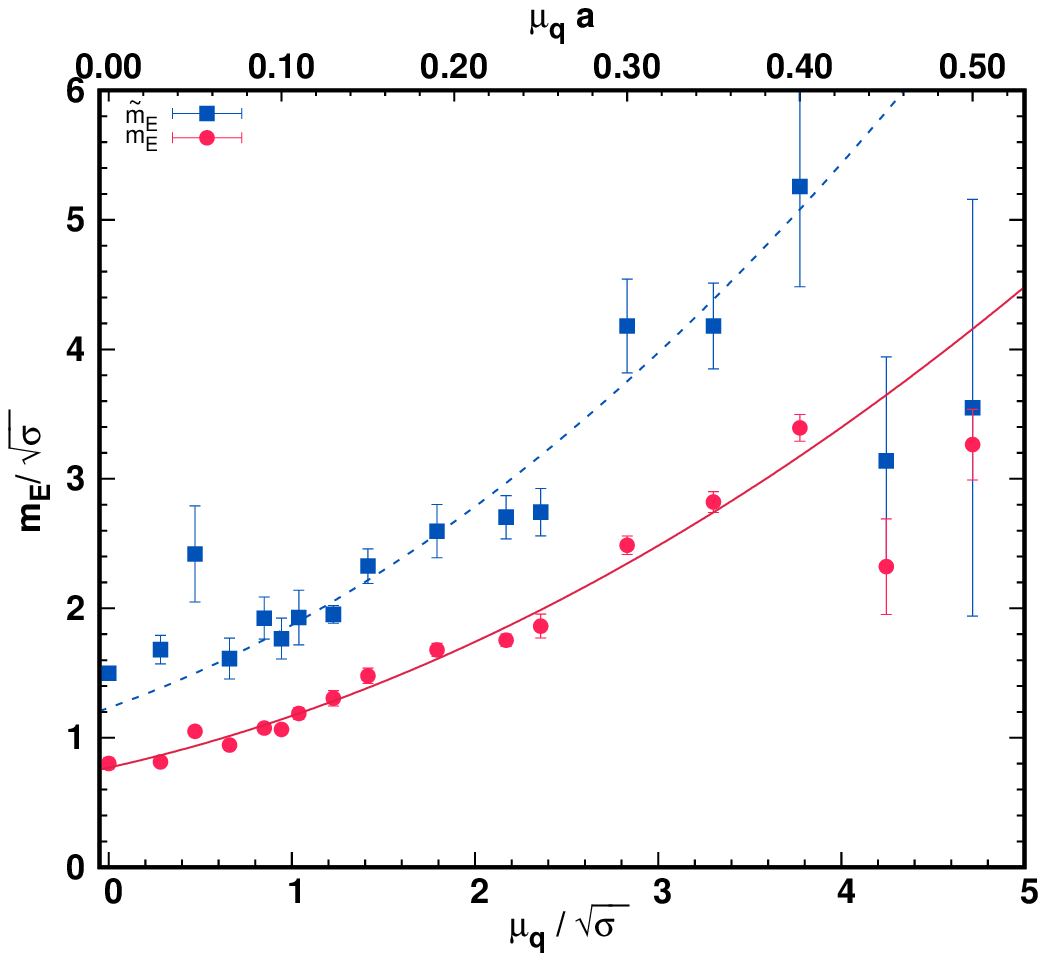}
\hspace*{-15mm}
\includegraphics[width=9cm]{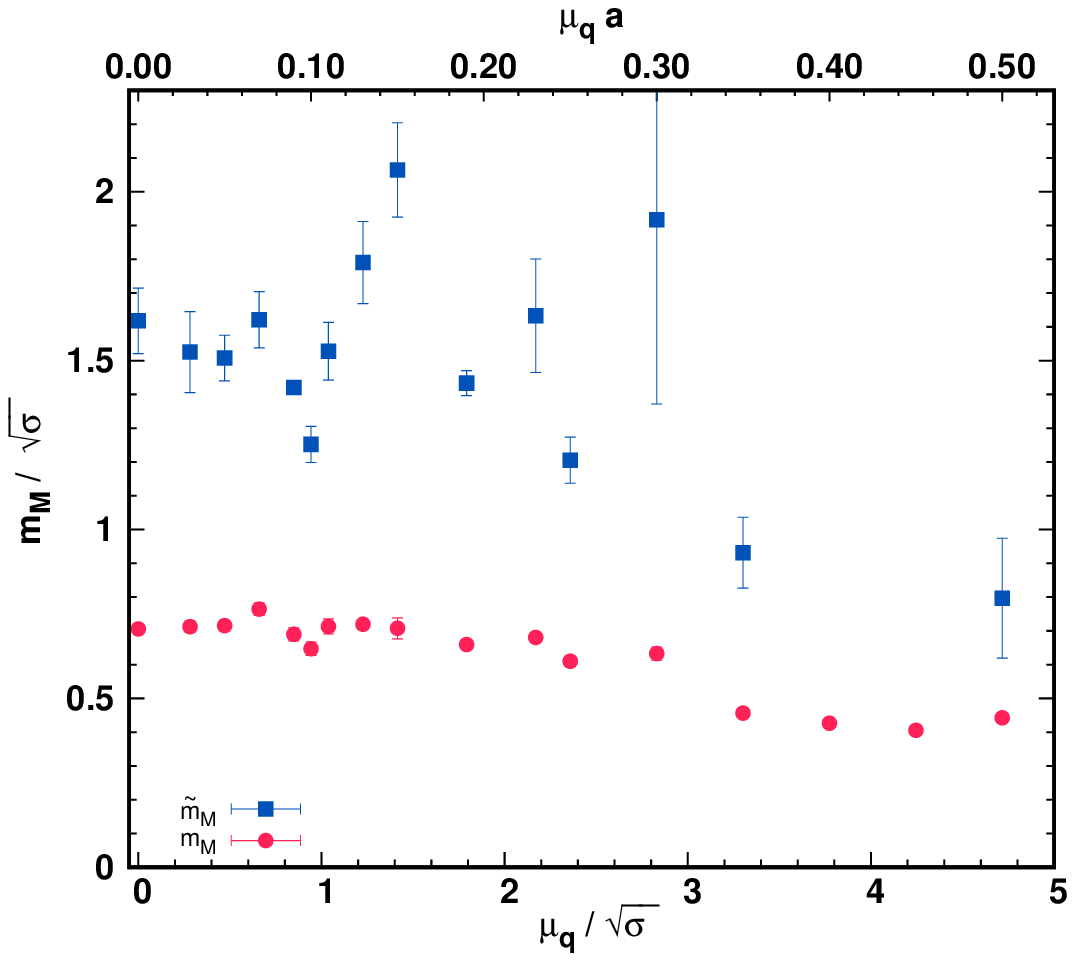}
\vspace*{0mm}
\caption{Electric (left panel) and magnetic (right panel) screening masses defined by
eq.~(\ref{mass_def1}-\ref{eq:correlation}) (squares) and by eq.~(\ref{maas}) (circles)
as functions of $\mu_q$. The curves in the left panel are described in the text.)
}
\label{fig:mass_vs_mu}
\end{figure}

In Fig.\ref{fig:mass_vs_mu} we show the electric (left panel) and magnetic (right panel) masses defined
according to these two definitions.
Our value for $\tilde{m}_E/\sqrt{\sigma}$ at $\mu_q=0$ is 1.50(4). This value can be
compared with the value 1.47(2)
obtained in SU(3) gluodynamics at zero temperature \cite{Oliveira:2010xc}   by fitting the
inverse propagator to the form  (\ref{scrmass_mom1})  at small momenta\footnote{We obtained this value taking
mass value 647(7) MeV, obtained in \cite{Oliveira:2010xc} and dividing it by $\sqrt{\sigma} = 440$MeV used
in \cite{Oliveira:2010xc} to set the scale.} We also quote a value  1.48(5) obtained for a mass dominating the small momentum behavior of a gluon propagator in  $SU(2)$ lattice gluodynamics in  \cite{Langfeld:2001cz}.

One can see that $m_E$ and $\tilde{m}_E$ show qualitatively very similar dependence
on $\mu_q$. They do not change much at small $\mu_q$ corresponding to the hadron phase.
Above $\mu_q/\sqrt{\sigma} \approx 0.5$ they start to increase and continue to increase at all $\mu_q$ values.
This behavior is similar to increasing of the electric screening mass with increasing temperature
in QCD at $T>T_c$ as was demonstrated by lattice simulations with definition (\ref{maas}) in
\cite{Fischer:2010fx,Maas:2011ez,Silva:2013maa} and with definition (\ref{mass_def1}-\ref{eq:correlation})
in \cite{Bornyakov:2010nc,Silva:2013maa}.
No such increasing was reported in Ref.~\cite{Boz:2018crd}.

In Ref.~\cite{Bornyakov:2019jfz} we found that the ratio $\tilde{m}_E/m_E$  can be well  approximated by a constant 1.6 for the range $0.9 < \mu_q/\sqrt{\sigma} < 3.0$. Now we can confirm this conclusion for larger $\mu_q$ included in this paper. The lower curve in this Figure shows fit of $m_E$ values by a polynomial of degree two. The upper curve is obtained by multiplication with factor 1.6. One can see that the upper curve agree well with $\tilde{m}_E$. The visible deviation is observed for the hadron phase only as we reported in Ref.~\cite{Bornyakov:2019jfz}. 

From Fig.\ref{fig:mass_vs_mu} (right) one can see that the magnetic screening masses $m_M$ and $\tilde{m}_M$  also  have qualitatively similar dependence on  $\mu_q$, although with one exception:
$\tilde{m}_M$  shows increasing in the range  $ 1.0 \lesssim \mu_q/\sqrt{\sigma} \lesssim 1.5.$ while 
$m_M$ is not increasing. 
Further, Fig.\ref{fig:mass_vs_mu} shows that for $\mu_q/\sqrt{\sigma} \gtrsim 3.4$ 
the values of both $\tilde{m}_M$ and $m_M$ are 
smaller than their values at lower $\mu_q$. Thus, we find an indication that the magnetic screening length is increasing at
large chemical potential in opposite to the electric screening length and in agreement with perturbation theory. No similar decreasing of $m_M$ was observed in the high temperature QCD or high temperature gluodynamics.
Note, that the range of $\mu_q/\sqrt{\sigma} \gtrsim 3.4$ is roughly corresponding to the range
where the spatial string tension $\sigma_s$ is zero, see Fig.5 in Ref.~\cite{Bornyakov:2017txe}.    

Comparing with results of Ref.~\cite{Boz:2018crd} we note that the fluctuation of $m_M$
around a constant value at smaller values of $\mu_q$ was also observed in that paper.
At large values of $\mu_q$ no decreasing of $m_M$ was found in Ref.~\cite{Boz:2018crd}.
In opposite, the results of Ref.~\cite{Boz:2018crd} hint to increasing of $m_M$ at large $\mu_q$.

\begin{figure}[tbh]
\begin{center}
\includegraphics[width=11cm]{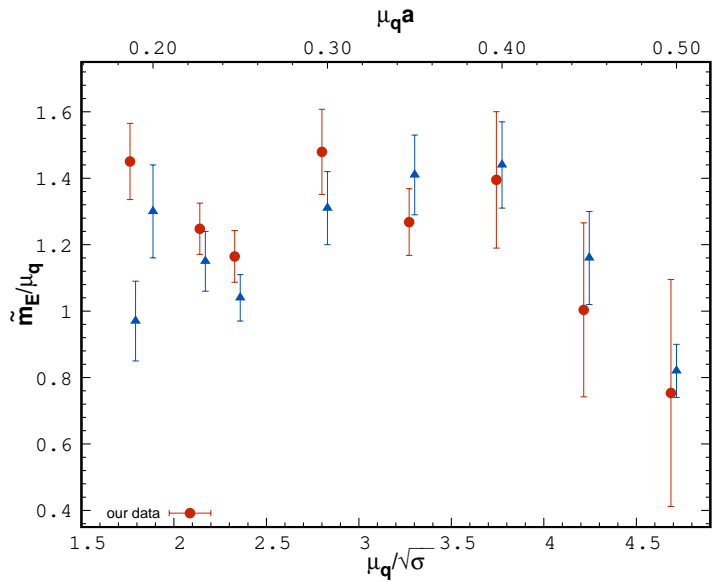}
\vspace*{0mm}
\end{center}
\caption{Comparison of the electric screening masses $\tilde{m}_E$ and Debye mass $m_D$ computed in Ref.~\cite{Astrakhantsev:2018uzd}
}
\label{fig:mass_comparison}
\end{figure}

In Ref.~\cite{Astrakhantsev:2018uzd} we computed the Debye screening mass $m_D$ from the
singlet quark-antiquark potential at large distances using the Coulomb gauge. It is expected that $m_D$ should agree with the electric screening mass computed from the gluon propagator. In Fig.\ref{fig:mass_comparison} we compare
$\tilde{m}_E$ and $m_D$. One can see the agreement within a standard deviation at all values of $\mu_q$
in the deconfinement phase, i.e. at $\mu_q/\sqrt{\sigma} > 1.9$. 
Thus, the values of the electric screening mass
computed using two different approaches in two different gauges coincide over a wide range of $\mu_q$.
We consider this as an important result because it gives some evidence for gauge invariance of the electric screening mass.
Note also that the ratio $\tilde{m}_E/\mu_q$ is a slowly varying function of $\mu_q$ in a qualitative agreement 
with perturbation theory.

We end this section with a remark on the reason for the differences between our results for the screening masses and results of Ref.~\cite{Boz:2018crd}. We use a very small value of the lattice spacing in our simulations.
This allows us to reach large physical values of $\mu_q$ 
keeping $a\mu_q$ small.
In opposite, the values of lattice spacing used in Ref.~\cite{Boz:2018crd}) are at least three times greater and this might cause large lattice artifacts at large $\mu_q$. 
Another source of the difference in results is the difference in the fermion action discretization
used in this paper and in Ref.~\cite{Boz:2018crd}). Thus results with the Wilson fermions and
small lattice spacing are highly needed.

\section{$D_L-D_T$ as an indicator of transitions}
\label{section4}

In the previous two sections we demonstrated that the propagators $D_L(p)$ and $D_T(p)$ become
more and more different in the infrared region when the chemical potential is increasing.
At the same time they approach each other at high momenta for fixed $\mu_q$.
In this section we study how fast they approach each other with increasing momentum and how
the picture changes with increasing $\mu_q$.
Similar comparison of these two propagators was made in Ref.~\cite{Silva:2013maa}
in finite-temperature $SU(3)$ gluodynamics where their ratio computed.
It was demonstrated that $D_L(p)$ dominates over $D_T(p)$ in the confinement phase at all momenta, whereas  $D_T(p)$ becomes dominating at high enough momenta in the deconfinement phase.

We show below that, in the theory under study, the difference between the transverse and
longitudinal propagators, $\Delta(p)=D_T(p)-D_L(p)$ has interesting dependence both on
momentum and on chemical potential.
The important finding is that the soft mode  $\Delta(p), p_4=0$ which is studied here
shows clear exponential dependence on $p$, which was observed recently  also in $SU(2)$ gluodynamics at finite temperatures \cite{BornyakovRogalyov2020}.

\begin{figure}[tbh]
\begin{center}
\vspace*{0mm}
\includegraphics[width=9cm]{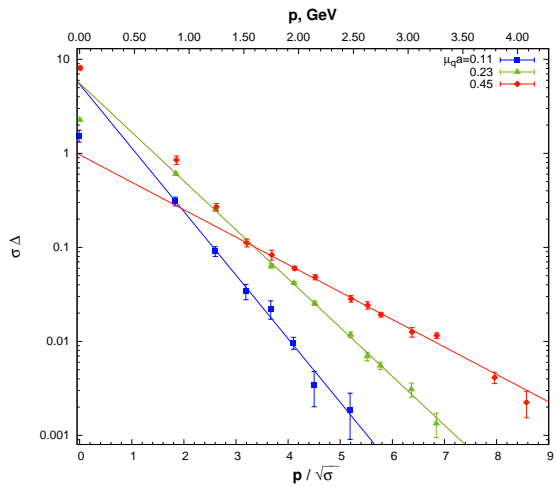}
\vspace*{0mm}
\end{center}
\caption{Difference $D_T-D_L$ as functions of $p$
at few values of $\mu_q$.
}
\label{fig:dglp_vs_p}
\end{figure}

Our numerical results for $\Delta(p)$  are presented in Fig.\ref{fig:dglp_vs_p}.
We show data at $\mu_q/\sqrt{\sigma}=1.0, 2.2$, and 4.2.
The exponential decreasing is well established starting from some momentum $p_0$ depending on $\mu_q$.
We found that $p_0=p_{min}$ for $ 1.0 \leq  \mu_q/\sqrt{\sigma}  \leq 3.0$ and $p_0/\sqrt{\sigma} \approx 3.2$ for higher $\mu_q$.

Thus we arrive at a simple fit function to describe the momentum dependence of $\Delta(p)$ at $p>p_0$.
\beq \label{eq:expo_fit_for_Delta}
\Delta(p) = c \exp (- \nu \cdot p)\ ,
\eeq
As a check we compared the fit by function (\ref{eq:expo_fit_for_Delta})  with the fit
by function
\beq\label{eq:pow_fit_for_Delta}
\Delta(p) = d \cdot p^{v}\,
\eeq
motivated by a power-like behavior of both gluon propagators when $p \to \infty$.

We cannot perform fitting for $\mu_q/\sqrt{\sigma} <0.5$ since
$\Delta(p)$ differs from zero at two values of the momentum only.
For $0.5\leq \mu_q/\sqrt{\sigma} \leq 1.0$ $\Delta(p)$ does not vanish at a very few
momenta. For this reason, both fit functions work well.
At $\mu_q/\sqrt{\sigma} > 1.0 $ only the  fit function (\ref{eq:expo_fit_for_Delta})  works.
We show the results of our fits for this range of $\mu_q$ in Fig.\ref{fig:dglp_vs_p}.

\begin{figure}[htb]
\hspace*{-5mm}
\includegraphics[width=8.5cm]{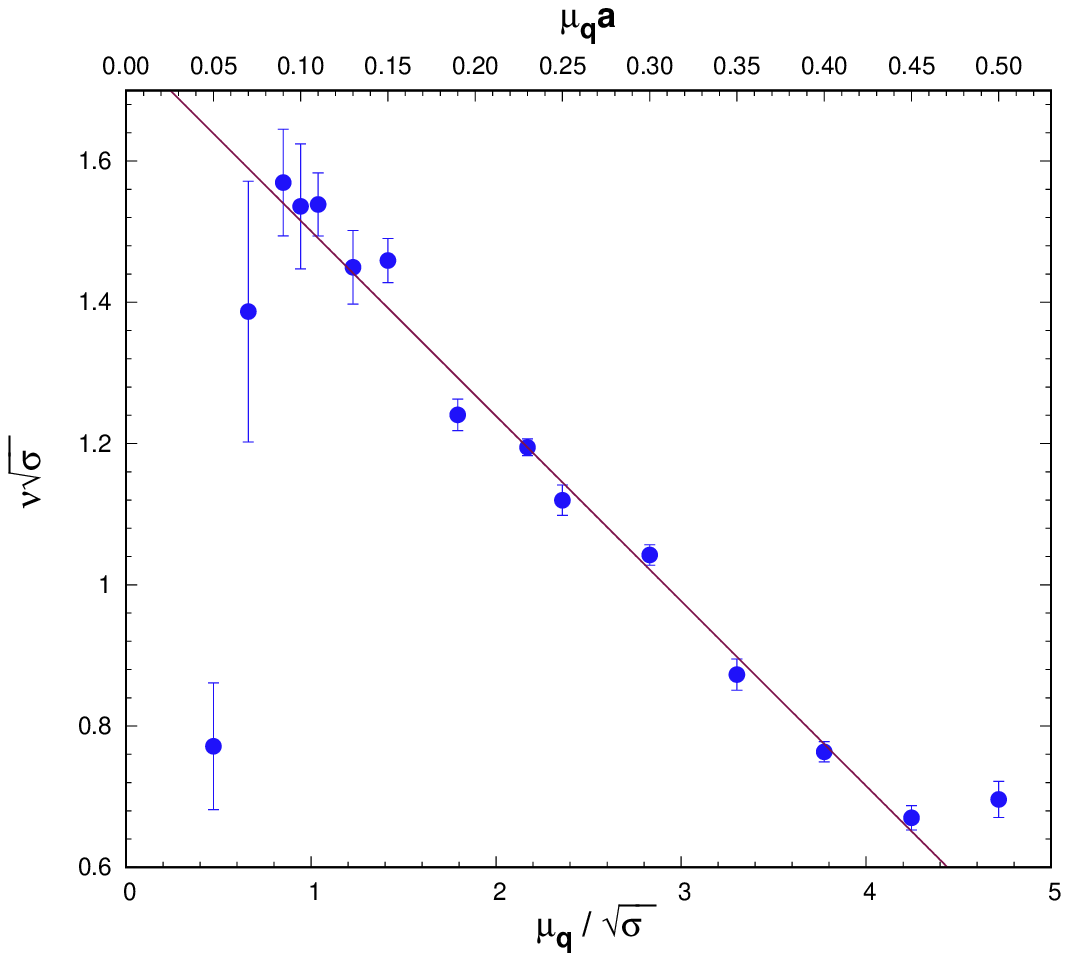}
\hspace*{-5mm}
\includegraphics[width=8.5cm]{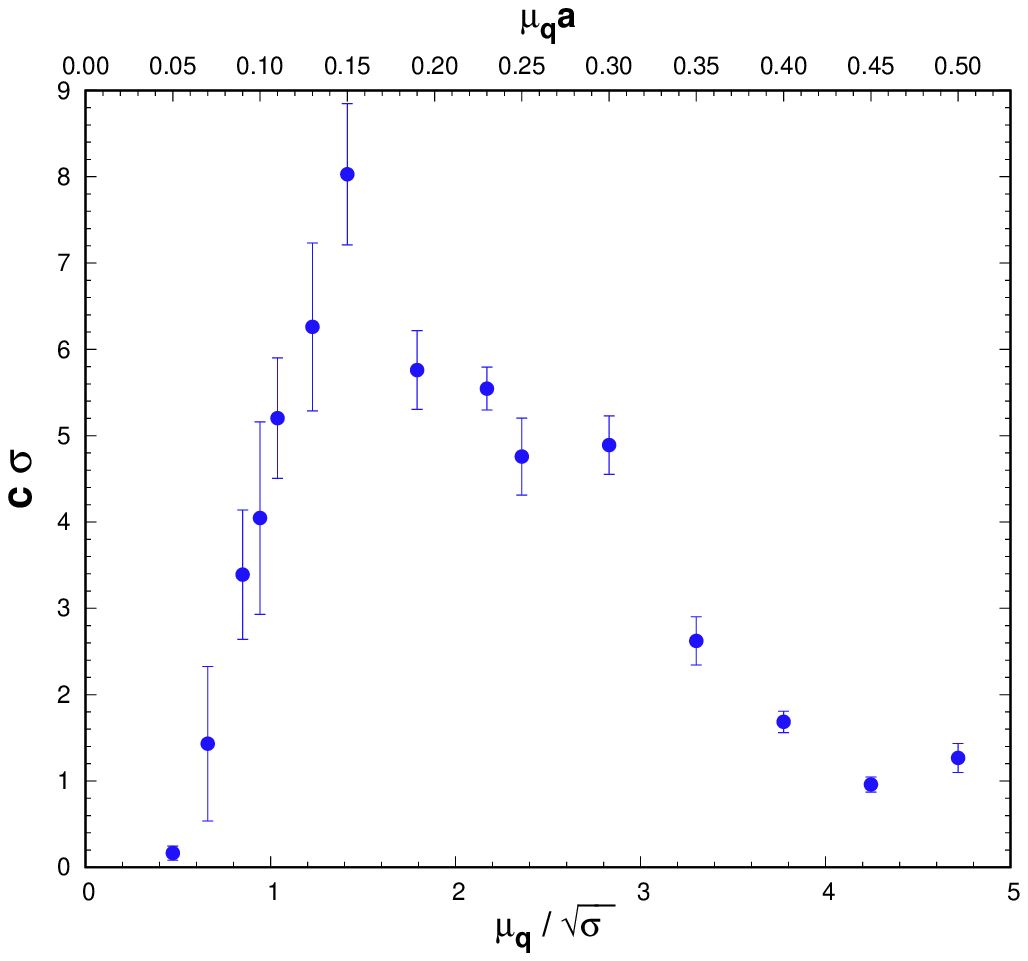}
\vspace*{0mm}
\caption{Parameters of the fit  (\ref{eq:expo_fit_for_Delta}) as functions of $\mu_q$.
}
\label{fig:exp_param_vs_mu}
\end{figure}

The dependence of the parameters $c$ and $\nu$ on the quark chemical potential
is shown in Fig.~\ref{fig:exp_param_vs_mu}.
The exponent $\nu$ is linearly decreasing over the range $ 1.0 \leq \mu_q/\sqrt{\sigma}  \leq 4.2$:
$\nu(\mu_q)$ can be fitted by the linear function
\beq
\nu = \nu_0 - \nu_1 \mu_q \;,
\eeq
where $\nu_0=1.76(3)/\sqrt{\sigma}$, $\nu_1=0.26(1)/\sigma$,
$\ds {\chi^2\over N_{d.o.f}}=2.0$ ($p-$value~$=0.04$).

\section{Conclusions}
\label{section5}
We presented results of our study of the longitudinal and transverse propagators
in the Landau gauge of the QC$_2$D with $N_f=2$ lattice staggered quark action at nonzero quark chemical potential. In contrast to earlier studies of the gluon propagators in this theory
\cite{Hands:2006ve,Boz:2013rca,Hajizadeh:2017ewa,Boz:2018crd}, we employed lattices with a very small lattice spacing and thus reached large physical values of $\mu_q$ keeping lattice values of $a\mu_q$ small.

We checked the effects of Gribov copies and found no such  effects even in the infrared region. 
This is different from the results of lattice gluodynamics. There are two reasons for this difference. The $Z_2$ center symmetry which is a source of the Gribov copies in the gluodynamics with 
periodic boundary conditions is broken in a theory with the matter field. 
Another reason is that the physical volume of our lattices is rather small.

Our main observations are as follows. We found that the longitudinal propagator
$D_L(p)$ is more and more suppressed in the infrared with increasing $\mu_q$.
This is reflected in particular in the increasing of the electric screening mass.
Such dependence of $D_L(p)$ on $\mu_q$ is analogous to its dependence on the temperature at $T>T_c$.
In opposite, we found much  weaker dependence on $\mu_q$ for the transverse propagator 
$D_T(p)$ with indication of the infrared enhancement at large $\mu_q$. 

We considered two definitions of the screening mass. The definition eq.~(\ref{maas}) is widely used though it has some drawbacks, in particular it depends on renormalization. The other definition eq.~(\ref{eq:LOW_MOM_fit_fun}) is renormgroup invariant. We found that both electric masses increase with $\mu_q$ and their ratio is 
a constant factor. A similar relation between the magnetic masses $\tilde{m}_M$ and $m_M$ is not ruled out although our results for $\tilde{m}_M$ have rather large statistical errors.

It is encouraging that our value  $\tilde{m}_E/\sqrt{\sigma} = 1.50(4)$ obtained at
$\mu_q=0$ is in a good agreement with respective values found in $SU(2)$ \cite{Langfeld:2001cz} and $SU(3)$ \cite{Oliveira:2010xc} lattice gluodynamics.

Another important result concerning the electric screening mass is a very good agreement between  $\tilde{m}_E$ and the Debye screening mass $m_D$ determined
from the singlet quark-anti-quark potential at large distances, see Fig.~\ref{fig:mass_comparison}.
This result indicates gauge invariance of the electric screening mass~(\ref{eq:LOW_MOM_fit_fun}).

For the magnetic screening masses we found that they show only a weak dependence on $\mu_q$ at $\mu_q \lesssim 2.2 \sqrt{\sigma} $ with clearly lower values at $\mu_q \gtrsim 3.4 \sqrt{\sigma}$. 
As we know from our previous study \cite{Bornyakov:2017txe}, this is the range where the spatial string tension becomes zero.
This decreasing of the magnetic screening mass is also in agreement with disappearance of the magnetic field screening at extremely large quark chemical potential predicted in \cite{Son:1998uk}.

Both increasing of the electric screening mass and decreasing of 
the magnetic screening mass at high quark densities were not observed
before in simulations with Wilson fermions on coarse lattices~\cite{Hands:2006ve,Boz:2013rca,Hajizadeh:2017ewa,Boz:2018crd}.

We also  studied the difference $\Delta(p) = D_L(p)-D_T(p)$ and found that it decreases
exponentially with momentum at large $p$. The respective exponent is decreasing
linearly with $\mu_q$ thus indicating that asymmetry between the propagators survives for higher momenta with increasing $\mu_q$.

\acknowledgments
The work was completed due to support of the Russian Foundation for Basic Research via grant
18-02-40130 mega. V.~V.~B. acknowledges the support from the BASIS foundation. A. A. N. acknowledges the support from STFC via grant
ST/P00055X/1. The authors are thankful to Andrey Kotov for participation in the project at the earlier stage and to Jon-Ivar Skullerud and  Etsuko Itou for useful discussions.

The research is carried out using the Central Linux Cluster of the NRC
"Kurchatov Institute" - IHEP,  the equipment of the shared research facilities of HPC computing resources at Lomonosov Moscow State University, the Linux Cluster of the NRC "Kurchatov Institute"
- ITEP (Moscow).  In addition, we used computer resources of the federal
collective usage center Complex for Simulation and Data Processing for
Mega-science Facilities at NRC “Kurchatov Institute”, http://ckp.nrcki.ru/.

\newpage

\appendix
\section{Fit results}

\begin{table}[h]
\begin{center}
\vspace*{0.5cm}
\begin{tabular}{|c|c|c|c|c|c|c|} \hline
~$\mu_q/\sqrt{\sigma}$~&~$M_L/\sigma$ &~~$R_L/\sigma$~~&~~$\delta_L \sigma $~~&  p-value & $\chi^2/N_{dof}$ & $Z_L/\sigma$ \\ \hline\hline
0.00  &  3.34(17)  &  2.12(48)  & 0.060(8)   &  0.68  & 0.67 & 16.9 \\
0.28  &  2.90(13)  &  0.95(25)  & 0.082(9)   &  0.65  & 0.54 & 12.6  \\
0.47  &  3.82(17)  &  0.79(24)  & 0.076(8)   &  0.66  & 0.53 & 13.2  \\
0.66  &  3.80(30)  &  1.91(67)  & 0.064(13)  &  0.27  & 1.30 & 15.8  \\
0.85  &  4.45(33)  &  1.89(61)  & 0.057(11)  &  0.46  & 0.86 & 17.0  \\
0.94  &  4.80(71)  &  3.2(1.6)  & 0.050(19)  &  0.04  & 2.48 & 19.9  \\
1.04  &  5.31(36)  &  2.95(74)  & 0.049(9)   &  0.65  & 0.67 & 20.0  \\
1.23  &  8.0(1.2)  &  7.9(3.5)  & 0.023(11)  &  0.17  & 1.49 & 37.6  \\
1.42  &  8.01(49)  &  5.6(1.2)  & 0.031(6)   &  0.95  & 0.35 &  29.4  \\
1.79  & 10.1(1.4)  &  7.7(2.8)  & 0.024(10)  &  0.38  & 1.08 & 37.0  \\
2.17  & 10.0(1.0)  &  6.4(1.9)  & 0.028(8)   &  0.11  & 1.58 & 31.8  \\
2.36  & 12.8(1.1)  & 10.2(2.9)  & 0.019(5)   &  0.56  & 0.88 & 44.4 \\
2.83  & 18.4(2.3)  & 13.3(3.9)  & 0.017(6)   &  0.49  & 0.95 & 53.0   \\
3.77  & 22.6(4.0)  & 14.5(4.7)  & 0.022(10)  &  0.63  & 0.82 & 45.8  \\
\hline\hline
\end{tabular}
\end{center}
\caption{Parameters of the  fits of $D_L(p)$ to function (\ref{eq:GS_fit_widely_used}).
}
\label{tab:LOW_MOM_fit_fun_L}
\end{table}
\begin{table}[h]
\begin{center}
\vspace*{0.5cm}
\begin{tabular}{|c|c|c|c|c|c|c|} \hline
 ~$\mu_q /\sqrt{\sigma}$~ & ~$M_T/\sigma$~ & ~$R_T/\sigma$~ &  ~$\delta_T \sigma$~ & $p-$value & ~$\chi^2 / N_{dof}$ & $Z_T/\sigma$ \\ \hline\hline
0.00  &  3.26(26) &  2.93(79) &  0.044(9) &   0.05  &  1.94 &  21.5  \\
0.28  &  3.09(21) &  2.49(60) &  0.052(8) &   0.00  &  3.09 &  19.1 \\
0.47  &  3.11(14) &  2.45(39) &  0.052(6) &   0.37  &  1.08 &  19.0  \\
0.66  &  3.32(20) &  2.52(54) &  0.052(8) &   0.24  &  1.29 & 19.0   \\
0.85  &  2.91(11) &  1.92(30) &  0.056(5) &   0.51  &  0.90 &  17.6 \\
0.94  &  2.57(15) &  1.80(41) &  0.067(8) &   0.38  &  1.07 &  15.6  \\
1.04  &  2.78(15) &  1.26(35) &  0.068(8) &   0.12  &  1.58 &  15.0  \\
1.23  &  2.66(13) &  0.92(26) &  0.076(8) &   0.14  &  1.53 &  13.6  \\
1.42  &  2.70(17) &  0.67(33) &  0.070(9) &   0.05  &  1.91 &  14.4  \\
1.79  &  3.39(11) &  2.40(31) &  0.032(3) &   0.50  &  0.92 &  26.3  \\
2.17  &  3.86(19) &  3.14(59) &  0.024(4) &   0.08  &  1.75 &  32.1  \\
2.36  &  3.51(19) &  3.44(66) &  0.024(4) &   0.53  &  0.88 &  32.9  \\
2.83  &  3.26(17) &  1.68(47) &  0.032(4) &   0.37  &  1.09 &  26.7  \\
\hline\hline
\end{tabular}
\end{center}
\caption{Parameters of the  fit of $D_T(p)$ to function (\ref{eq:GS_fit_widely_used}).
}
\label{tab:LOW_MOM_fit_fun_T}
\end{table}

\end{document}